\documentclass[preprint]{elsarticle}

\usepackage[margin=2.5cm]{geometry}
\usepackage{amsmath,amsfonts}
\usepackage{bbding}
\usepackage{array}
\usepackage[caption=false]{subfig}
\usepackage{textcomp}
\usepackage{url}
\usepackage{verbatim}
\usepackage{graphicx}
\usepackage{tabularx}
\usepackage{booktabs}
\usepackage{array}
\usepackage{multirow}
\usepackage{xcolor}
\usepackage[hidelinks]{hyperref}
\usepackage{cleveref}
\usepackage{mathtools}
\usepackage{amsmath}
\usepackage{amssymb}
\usepackage{mathrsfs}

\crefname{equation}{}{}
\Crefname{equation}{Equation}{Equations}
\crefname{figure}{Fig.}{Figs.}
\crefname{table}{Table}{Tables}
\crefname{section}{Section}{Sections}
\crefname{algorithm}{Algorithm}{Algorithms}

\begin{document}

\makeatletter
\def\@date{}
\def\ps@pprintTitle{%
 \let\@oddhead\@empty
 \let\@evenhead\@empty
 \let\@oddfoot\@empty
 \let\@evenfoot\@empty
}
\makeatother

\begin{frontmatter}

\title{Adaptive Informed Deep Neural Networks for Power Flow Analysis}

\author{Zeynab Kaseb\corref{correspondingauthor}}
\author{Stavros Orfanoudakis}
\author{Pedro P. Vergara}
\author{Peter Palensky}

\address{Department of Electrical Sustainable Energy, Delft University of Technology, The Netherlands}

\cortext[correspondingauthor]{Corresponding author: Zeynab Kaseb. Email: \texttt{Z.Kaseb@tudelft.nl}}

\begin{abstract}
This study introduces PINN4PF, an end-to-end deep learning architecture for power flow (PF) analysis that effectively captures the nonlinear dynamics of large-scale modern power systems. The proposed neural network (NN) architecture consists of two important advancements in the training pipeline: (A) a double-head feed-forward NN that aligns with PF analysis, including an activation function that adjusts to the net active and reactive power injections patterns, and (B) a physics-based loss function that partially incorporates power system topology information through a novel hidden function. The effectiveness of the proposed architecture is illustrated through 4-bus, 15-bus, 290-bus, and 2224-bus test systems and is evaluated against two baselines: a linear regression model (LR) and a black-box NN (MLP). The comparison is based on (i) generalization ability, (ii) robustness, (iii) impact of training dataset size on generalization ability, (iv) accuracy in approximating derived PF quantities (specifically line current, line active power, and line reactive power), and (v) scalability. Results demonstrate that PINN4PF outperforms both baselines across all test systems by up to two orders of magnitude not only in terms of direct criteria, e.g., generalization ability, but also in terms of approximating derived physical quantities.
\end{abstract}


\end{frontmatter}


\section{Introduction}
Power flow (PF) analysis is a foundational computational method to assess and determine the steady-state operating conditions of electrical power systems by computing voltage magnitudes and phase angles at all buses. This analysis is crucial to ensure the reliability, stability, and optimal performance of power systems. It also allows operators to make informed decisions and mitigate potential issues, such as voltage violations, overloads, and system instability \cite{Li2025,Buason2024AdaptiveInsights}.

PF analysis can be performed by solving nonlinear and non-convex algebraic equations derived from the nodal balance of the net active and reactive power injections at each bus in power systems \cite{Bugosen2024ApplicationsEquations}. Exact analytical solutions for these equations, which also involve impedance parameters, load characteristics, and generator conditions of power systems, are impossible. Therefore, iterative numerical methods, such as the Gauss-Seidel and Newton-Raphson (NR) methods, are conventionally employed to converge to a solution that satisfies the PF equations within specified accuracy limits. Eventually, the solutions yield voltage phasors across the entire power system and, hence, provide a comprehensive understanding of the operational state of the power system \cite{Wang2023NovelRegulation}.

Conventional iterative numerical methods, however, face computational challenges in large-scale modern power systems. They struggle to capture the intricacies of power systems due to uncertainties involved, such as inaccurate line profiles due to weather conditions and aging, different types of loads, but also missing data on renewable energy resources \cite{Kaseb2024PowerApproach}. Ineffective PF analysis under these circumstances can lead to safety threats, including renewable energy generation curtailment and blackouts, and difficulties in accommodating distributed energy resources \cite{Sharma2021MajorChallenges,Huo2024GraphMicrogrid}. Addressing these challenges necessitates developing new approaches for PF analysis that are both computationally efficient and numerically stable.

Deep learning approaches, and more specifically, artificial neural networks (NNs), are currently the most powerful set of numerical tools to provide accurate approximations of nonlinear problems (e.g., \cite{Dziugaite2017ComputingData,Wolgast2024,Cibaku2024}). Several studies have demonstrated the superiority of deep learning approaches in PF analysis in terms of computational time by orders of magnitude (e.g., \cite{Beinert2023,Wang2022,Ahshan2025}). At the same time, the accuracy of the solutions is competitive compared to the conventional iterative numerical methods (e.g., \cite{Huang2023ApplicationsReview,Kaseb2024QuantumAnalysis}). NNs, therefore, can address the challenges mentioned above by leveraging the availability of massive measurements and/or augmented data, learning complex input-output relationships that are often difficult or even impossible for conventional iterative numerical methods to comprehend, and achieving the accuracy required for real-world applications \cite{Oliveira2023DeepQuality}. Nevertheless, NNs are subject to overfitting, the lack of generalization, and scalability issues. They are very unlikely to meet the physical constraints. Moreover, their performance relies heavily on the training dataset size and quality. In contrast, not enough data is always available due to privacy reasons and the presence of missing data, among others \cite{vonRueden2019InformedSystems}.

Several studies in the literature have investigated the impact of various modifications on the efficacy of deep learning-based PF analysis (e.g., \cite{Li22025,Liang2025}). These modifications are categorized into three main stages: (i) pre-training (e.g., \cite{Lei2021Data-DrivenApproach}); (ii) training (e.g., \cite{Yang2020FastApproach}); and (iii) post-training (e.g., \cite{Donti2021DC3:Constraints}), as outlined in Table~\ref{tab:literature}. A majority of these studies have primarily focused on the training stage. The table also highlights instances where topology information, such as line physical properties (e.g., \cite{Jeddi2021AAnalysis}) and graph topology (e.g., \cite{DeJongh2022Physics-InformedSystems}) has been integrated into the training stage. The literature review also indicates the utilization of different prior knowledge for deep learning-based PF analysis and optimal power flow (OPF), including equality and inequality constraints (e.g., \cite{Yang2023Physics-GuidedAnalysis}) and PF equations (e.g., \cite{Kody2022ModelingCommitment}). Among the three presented stages, this work focuses on the training stage.


\begin{table}[t!]
\caption{List of literature on deep learning-based power flow (PF) and optimal power flow (OPF). Studies are grouped into: (i) pre-training; (ii) training; (iii) post-training.}
\label{tab:literature}
\centering
\scriptsize
\renewcommand{\arraystretch}{1.4}
\begin{tabularx}{\textwidth}{p{1.3cm} p{2.8cm} p{1.4cm} p{2cm} p{2cm} X}
\toprule
Stage & Study & Application & Input$^{*}$ & Output$^{*}$ & Approach \\[2pt]
\midrule

Pre-training & Lei et al. (2021) \cite{Lei2021Data-DrivenApproach} & OPF & $p^d$, $q^d$ & $p^g$, $q^g$, $|v_i|$, $\delta_i$  & Integrating measurement data into the pre-trained NN \\[2pt]

Training & Yang et al. (2020) \cite{Yang2020FastApproach} & PF & $p_{i}$, $q_{i}$ & $|v_i|$, $\delta_i$ & Adding a penalty term of branch flows to loss function\\[2pt]

& Jeddi \& Shafieezadeh (2021) \cite{Jeddi2021AAnalysis} & PF & $\Delta p_{i}$, $\Delta q_{i}$, $|v_{i}|$, $\delta_{i}$, $l_{ij}$ & $|v_{i}|$, $\delta_{i}$ & Using self-attention mechanism and adding a penalty term of nodal voltage to loss function\\[2pt]

& Hu et al. (2021) \cite{Hu2021Physics-GuidedAnalysis} & PF & $p^d$, $q^d$, $p^g$, $|v_{g}|$ & $|v_i|$, $\delta_i$ & Adding regularization terms based on structure of AC PF equations and topology of power system\\[2pt]

& De Jongh et al. (2022) \cite{DeJongh2022Physics-InformedSystems} & PF & $p_{i}$, $q_{i}$, $|v_i|$, $\delta_i$, $\mathcal{G}$ & $p_{i}$, $q_{i}$, $|v_i|$, $\delta_i$ & Adding a penalty term of nodal active and reactive power to loss function\\[2pt]

& Kody et al. (2022) \cite{Kody2022ModelingCommitment} & PF & $|v_i|$, $\delta_i$ & $p_{i}$, $q_{i}$, $s_{ij}$ & Adding a physics-based term to NN formulation\\[2pt]

& Yang et al. (2023) \cite{Yang2023Physics-GuidedAnalysis} & PF & $p_{i}$, $q_{i}$, $\mathcal{G}$, A & $|v_i|$, $\delta_i$ & Developing a topology-adaptive graph NN and adding penalty terms of equality constraints to loss function\\[2pt]

& Lin et al. (2023) \cite{Lin2024PowerFlowNet:Networks} & PF & $p^{d}$, $q^{d}$, $p^{g}$, $|v_g|$, $l_{ij}$, $\mathcal{G}$ & $p_{i}$, $q_{i}$, $|v_i|$, $\delta_i$ & Combining message passing graph NNs and high-order graph convolutional NNs\\[2pt]

& Liu et al. (2024) \cite{Liu2024VoltageData} & PF & $p^{d}$, $q^{d}$, $l_{ij}$ & $v_{i}$ &  Using a physics-inspired structure integrating physical laws of PF \\ [2pt]

& Nellikkath \& Chatzivasileiadis (2021) \cite{Nellikkath2021Physics-InformedFlow} & OPF & $p^d$ & $p^g$ & Adding a penalty term of active power generation and consumption to loss function\\[2pt]

& Nellikkath \& Chatzivasileiadis (2022) \cite{Nellikkath2022Physics-InformedFlow} & OPF & $p^d$, $q^d$ & $p^g$, $q^g$ & Adding a penalty term of active and reactive power generation and voltage magnitude to loss function\\[2pt]

& Hu \& Zhang (2023) \cite{Hu2023OptimalGeneration} & OPF & $p^d$, $q^d$ & $|v_i|$, $\delta_i$ & Developing an activation function and a modified loss function to satisfy physical constraints\\[2pt]

& Wu et al. (2024) \cite{Wu2024Physics-InformedResources} & OPF & $p_i$, $|v_i|$ & $|v_i|$, $\delta_i$ & Using a model-agnostic meta-learning algorithm and  enforcing PF equations as constraints in loss function\\[2pt]

Post-training & Donti et al. (2021) \cite{Donti2021DC3:Constraints} & OPF & $p^d$, $q^d$, $|v_r|$ & $p^g$, $|v_{r,g}|$  & Implementing completion idea (hard constraints)\\[2pt]

& Pan et al. (2022) \cite{Pan2022DeepOPF:Problems} & OPF & $p^d$, $q^d$ & $p^g$, $q^g$, $|v_i|$, $\delta_i$ & Implementing completion idea (hard constraints)\\[2pt]

& Li et al. (2023) \cite{Li2023Model-informedFlow} & OPF & $p^d$ & $p^g$, $\delta_i$ & Filtering generated values to ensure feasibility by comparing with actual values\\[2pt]

\bottomrule
\end{tabularx}

\noindent\footnotesize{$^{*}$$p_{i}$=nodal active power, $q_{i}$=nodal reactive power, $s_{ij}$=line apparent power, $|v|$=voltage magnitude, $\delta$=voltage angle, $\mathcal{G}$=graph topology, $l_{ij}$=line physical properties, $A$=Adjacency matrix, $p^{g}$=generator active power, $q^{g}$=generator reactive power, $p^{d}$=load active power, $q^{d}$=load reactive power.}
\end{table}

Measurement data has been used in \cite{Lei2021Data-DrivenApproach} to enhance deep learning-based OPF. While this modification has demonstrated improvements in accuracy, its applicability is confined to the pre-training stage. Physical properties have been used on a few occasions in the past to enhance deep learning-based PF analysis (e.g., \cite{Jeddi2021AAnalysis, Liao2024Data-drivenNetworks}), where the focus was mainly on the training stage. However, the high dependency on the physical properties of power systems affects the efficiency of this modification, certainly because physical properties are not always reliable due to different reasons, including aging and environmental conditions. In addition, integrating all the physical properties makes training processes computationally very expensive. The completion idea and filtering technique have also been employed in a few studies (e.g., \cite{Li2023Model-informedFlow}), where the focus is mainly on the post-training stage. Yet, similar to pre-training stage modifications, this approach has limitations as it is detached from the training stage. Among various methods, modified loss functions have been widely used for deep learning-based PF analysis (e.g., \cite{Hu2021Physics-GuidedAnalysis}) and OPF (e.g., \cite{Hu2023OptimalGeneration}). Physics-based loss functions have shown significant promise in enhancing the performance and reliability of NNs for PF and OPF.

This paper proposes an end-to-end deep learning architecture, hereafter called physics-informed neural network for power flow (PINN4PF), where the focus is on the training stage and extends the existing knowledge of physics-informed NNs for PF analysis (e.g., \cite{Yang2020FastApproach, Hu2021Physics-GuidedAnalysis}). PINN4PF learns the mapping from net active ($p$) and reactive ($q$) power injections to complex bus voltages. It employs a double-head feed-forward neural network, where one head predicts the real part ($\mu$) and the other the imaginary part ($\omega$) of the complex voltage. Each head uses a modified ReLU activation function with a trainable slope, enabling adaptive learning of nonlinear relationships during training via backpropagation. 

A key innovation is a physics-based loss function derived from the power balance equations, which incorporates prior knowledge of the system through a novel hidden function that partially encodes line characteristics. Unlike methods that require full system information (e.g., GNNs), our approach reduces dependency on complete topology data. Despite this, PINN4PF achieves up to two orders of magnitude improvement in performance compared to conventional NNs. The application of PINN4PF is illustrated through four test power systems, namely 4-bus \cite{Grainger1994PowerAnalysis}, 15-bus \cite{Farhangi2019MicrogridBenchmarks}, 290-bus \cite{Kerber2011AufnahmefahigkeitPhotovoltaikkleinanlagen}, and 2224-bus \cite{Bukhsh2013NetworkNetworks} test systems. The main contributions are:
\begin{enumerate}
    \item A double-head feed-forward NN for PF analysis, with adaptive activation functions responsive to the net active and reactive power injections patterns.
    \item A physics-based loss function that partially integrates power system topology, thus reducing the need for complete line data.
\end{enumerate}

\section{Power Flow Analysis}\label{sec:pf}
Power flow (PF) analysis aims to specify the state variables of power systems, i.e., $[\delta \, v]^T$, where $\delta$ and $v$ denote the voltage phase angle and magnitude, respectively. The AC PF analysis problem is a representation of Kirchhoff's laws and is formulated in rectangular coordinates \cite{Hu2021Physics-GuidedAnalysis} as:
\begin{subequations}
\begin{align}
    p_i
    &=
    \sum_{j=1}^{n} g_{ij}(\mu_i\mu_j+\omega_i\omega_j)+b_{ij}(\omega_i\mu_j-\mu_i\omega_j), \label{eq:pi}\\
    q_i
    &=
    \sum_{j=1}^{n} g_{ij}(\omega_i\mu_j-\mu_i\omega_j)-b_{ij}(\mu_i\mu_j+\omega_i\omega_j), \label{eq:qi}
\end{align}
\end{subequations}
where $i$ and $j$ are the indices of the buses, $n$ is the total number of buses in the power system, $p_i$ and $q_i$ are the net active and reactive power injections at bus $i$, $g_{ij}$ and $b_{ij}$ are the real and imaginary components of the admittance $y_{ij}$ between buses $i$ and $j$, $\mu_i=v_{i}cos\delta_{i}$ and $\omega_i=v_{i}sin\delta_{i}$ are the real and imaginary components of the voltage phasor at bus $i$. Here, $v_{i}$ and $\delta_{i}$ are the voltage magnitude and phase angle at bus $i$. 

There are three types of buses in power systems: (i) reference bus, (ii) load bus (\emph{pq} bus), and (iii) generation bus (\emph{pv} bus). $v_i$ and $\delta_i$ are known, and $p_i$ and $q_i$ are unknown for the reference bus. For load buses, $p_i$ and $q_i$ are known, and $v_i$ and $\delta_i$ are unknown. $v_i$ and $p_i$ are known for generation buses, while $q_i$ and $\delta_i$ are unknown. 

This study considers cases with one reference bus and load buses for simplicity. A standard approach to handling \emph{pv} buses is converting \emph{pv} buses to \emph{pq} buses by estimating the reactive power while applying constraints to maintain voltage levels within acceptable limits. Once converted, \emph{pv} buses can be treated as \emph{pq} buses for training. It should also be noted that while a voltage magnitude of $1.0$ p.u. and a phase angle of $0$ degrees are assumed in this study, it is flexible to accommodate different reference bus settings. 

For a power system consisting of a reference bus and load buses, a set of PF equations with the same number of equations and unknowns is achieved \cite{GoranAndersson2008ModellingPower}:
\begin{subequations} \label{eq:pq-mismatch}
\begin{align}
    p_i^d-p_i & = 0 \label{eq:p-mismatch},\\[2ex]
    q_i^d-q_i & = 0 \label{eq:q-mismatch},
\end{align}
\end{subequations}
where $p_i$ and $q_i$ are defined by \eqref{eq:pi} and \eqref{eq:qi}, respectively, and $p_{i}^{d}$ and $q_{i}^{d}$ are the net active and reactive power injections at bus $i$, respectively. \eqref{eq:pq-mismatch} is conventionally solved iteratively to specify $[\delta \, v]^T$ until a convergence criterion is met, i.e., the mismatch between $p_i$ and $p_{i}^{d}$ but also $q_i$ and $q_{i}^{d}$ are small enough.

\section{Deep Learning-based Power Flow Analysis}\label{sec:nn}
Deep learning-based PF analysis refers to developing NNs to approximate the state variables of power systems based on given historical system operation data, hereafter called \emph{dataset}. The dataset includes input features $\Vec{x}$ and output labels $\Vec{y}$. For a power system with a reference bus and load buses, the input features are known variables, i.e., the net active and reactive power injections at load buses $\vec{x} = \{(\vec{p}^d_i,\vec{q}^d_i): i=1,2,\dots,n\}$. The output labels are unknown variables, i.e., the real and imaginary components of complex voltages at load buses $\vec{y} = \{(\vec{\mu}_i,\vec{\omega}_i): i=1,2,\dots,n\}$. The dimension of $\vec{x}$ and $\vec{y}$ is therefore $n \times 2$, where $n$ is the number of load buses. Note that the voltage magnitude and phase angle are known for the reference bus $i=0$, and the net active and reactive power injections are unknown. Having voltages at all load buses approximated, the net active and reactive power injections at the reference bus can be calculated.

The training of NNs is an iterative process and involves four steps. In the first iteration, the set of trainable parameters of the NN, i.e., the weight matrices and bias vectors $\theta = \{ (W_k, b_k) : k=1,2,\dots,m \}$, are initialized in the \emph{update} step, where $m$ is the number of hidden layers. In the \emph{forward} step, the network is developed using \eqref{eq:functions-chain-1} and \eqref{eq:hidden_layer}. The deviations of the approximated output obtained by the NN $\hat{\vec{y}}$ from the output labels $\vec{y}$ are computed in the \emph{loss} step. Finally, the gradient of the deviations is calculated with respect to $\theta$ in the \emph{backward} step. For the next iteration, $\theta$ is fine-tuned in the \emph{update} step to reduce the deviations. The process continues until the maximum number of epochs is reached. In practice, the four steps can be individually and jointly modified to improve the overall performance of NNs. For example, it can be done by enhancing the network architecture in the \emph{forward} step or adding a physical penalty term to the loss function in the \emph{loss} step. 

\section{Proposed architecture: PINN4PF}\label{sec:pinn4pf}
\begin{figure*}[t]
\centering
\includegraphics[width=6.2in]{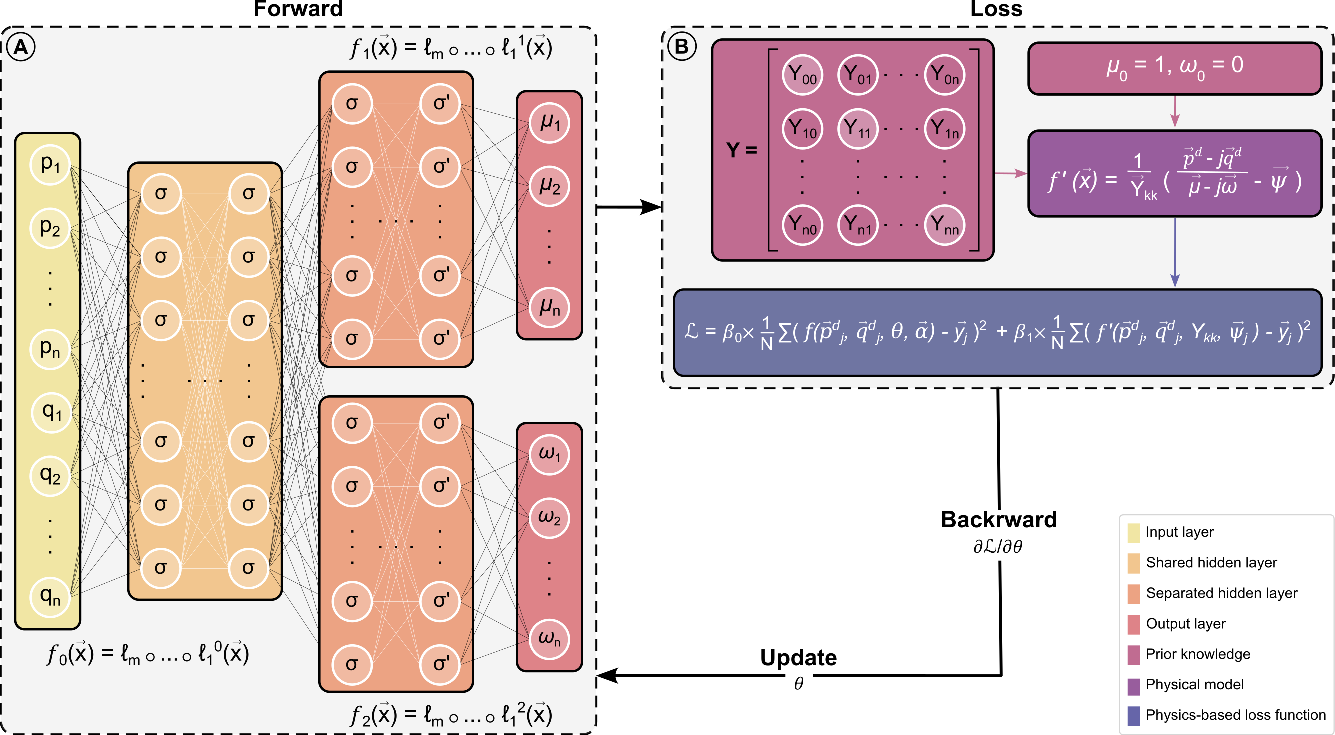}
\caption{Schematic diagram of the proposed PINN4PF architecture for PF analysis. The model incorporates two key advancements: (A) a double-head feed-forward NN, \( f(\cdot) \in \{f_0, f_1, f_2\} \), utilizing an adaptive activation function \( \sigma' (z) = \max(0, \alpha z) \) with a trainable slope \( \alpha \); and (B) a physics-informed loss function that combines supervised learning with physical constraints. The input layer receives a concatenated vector of active (\( \vec{p}^d \)) and reactive (\( \vec{q}^d \)) power at all \( n \) load buses, resulting in \( 2n \) input neurons. Shared hidden layers \( f_0(\cdot) \) act as a feature extractor, projecting the input to a higher-dimensional latent space. Two separate network heads, \( f_1(\cdot) \) and \( f_2(\cdot) \), then predict the real (\( \vec{\mu} \)) and imaginary (\( \vec{\omega} \)) components of the complex bus voltages, each with \( n \) output neurons. The model is trained by minimizing a composite loss that enforces data fidelity (using labels) and physical consistency to enable accurate and generalizable voltage predictions.}
\label{fig:framework}
\end{figure*}

The proposed physics-informed neural network for power flow (PINN4PF) includes two advancements in \emph{forward} and \emph{loss} steps, respectively denoted by A and B in Fig. \ref{fig:framework}, resulting in a double-head architecture enhanced with an adaptive activation function and a physics-based loss function. Note that classical PF solvers, e.g., the Newton-Raphson (NR) method, and PINN4PF can be interchangeably used to specify the state of the power system. The following subsections provide detailed information on the proposed novel advancements. 

\subsection{Forward Pass}\label{subsec:forward}
A double-head feed-forward NN $f(\cdot) \in \{(f_0, f_1, f_2)\}$ is developed to approximate $\hat{\vec{y}} = \{(\hat{\vec{\mu}}_i,\hat{\vec{\omega}}_i): i=1,2,\dots,n\}$ at load buses using the dataset $\{\vec{x}, \vec{y}\}$. It has three types of layers, i.e., input, hidden, and output layers, as highlighted in Fig. \ref{fig:framework}-A. The input and output layers correspond to the input features $\Vec{x}$ and output labels $\Vec{y}$, respectively. The neurons of the input layer contain active power $\vec{p}^d=[p_1,p_2,\dots,p_n]$ followed by reactive power $\vec{q}^d=[q_1,q_2,\dots,q_n]$ at all load buses, and hence, the input layer has $n \times 2$ neurons, where $n$ is the number of load buses. Following the input layer, there is a set of shared hidden layers $f_0(\cdot)$ acting as a feature extractor that projects $\vec{x}$ to a higher-dimensional space, where the two heads, $f_1(\cdot)$ and $f_2(\cdot)$, separately involve a few hidden layers to respectively approximate $\vec{\mu}=[\mu_1,\mu_2,\dots,\mu_n]$ and $\vec{\omega}=[\omega_1,\omega_2,\dots,\omega_n]$. Thus, the output layer of each head has $n$ neurons. The shared set of hidden layers and the two heads each are a chain of functions and can be represented as:
\begin{subequations}
\begin{align}
    f_0(\vec{x}) = l_m^0 \circ \ldots \circ l_1^0(\vec{x}), \label{eq:functions-chain-0} \\
    f_1(\vec{x}) = l_m^1 \circ \ldots \circ l_1^1(\vec{x}), \label{eq:functions-chain-1} \\
    f_2(\vec{x}) = l_m^2 \circ \ldots \circ l_1^2(\vec{x}), \label{eq:functions-chain-2} 
\end{align}
\end{subequations}
which sequentially process $\vec{x}$ through $m$ hidden layers to obtain $\vec{f}_0(\cdot)$, $\vec{f}_1(\cdot)$, and $\vec{f}_2(\cdot)$, respectively. For the shared set of hidden layers and the two heads, the $k$-th hidden layer ($k=1,2,\dots,m$) is given by:
\begin{align}
    l_{k}(\vec{x})=\sigma\left(W_{k}^{T} \cdot \vec{x} + b_k\right). \label{eq:hidden_layer}
\end{align}
Here, $W_{k}^{T}$ and $b_k$ are the weight matrix and bias vector for the corresponding hidden layer. Each hidden layer applies a linear transformation $W_{k}^{T} \cdot \vec{x} + b_k$, followed by a nonlinear transformation $\sigma(\cdot)$ to capture complex relationships between $\vec{x}$ and $\vec{y}$. Note that $W_k$ and $b_k$ are trainable parameters and are optimized during the training process. 

The Rectified Linear Unit (ReLU) activation function $\sigma(\cdot)$ is applied to the shared set of hidden layers of the two heads \eqref{eq:functions-chain-0}. While an adaptive version of ReLU $\sigma' (\cdot)$ is developed to apply model-based nonlinear transformations to the two heads \eqref{eq:functions-chain-1} and \eqref{eq:functions-chain-2}. That is, a trainable parameter $\alpha$ is introduced to scale the result of the linear transformation while applying nonlinearity. Hence, $\alpha$ serves as a guide to prevent overfitting and improve the generalization ability of the NN against unseen data. $\sigma(\cdot)$ and $\sigma'(\cdot)$ are represented as:
\begin{subequations}
\begin{align}
    \sigma = \max(0, z), \label{eq:relu}\\
    \sigma' = \max(0, \alpha z), \label{eq:adaptive-relu}
\end{align}
\end{subequations}
where $z$ is equivalent to $W_{k}^{T} \cdot \vec{x} + b_k$. During each iteration of the training process, the gradients of the loss function concerning the trainable parameters $\nabla_{W_k,b_k,\alpha_k} L$ are computed for each hidden layer $k$ using the chain rule of differentiation. The resulting gradients are then backpropagated through the network to update and optimize the weights $W_k$, biases $b_k$, and the scaling parameter $\alpha_k$. This process enables the network to learn the optimal values of $\alpha_k$, along with $W_k$ and $b_k$, and hence, potentially leads to improved performance in approximating $\hat{\vec{y}}$. More detailed information about adaptive activation functions can be found in \cite{KasebZ2023AdaptiveAnalysis, Jagtap2020AdaptiveNetworks}.

\subsection{Loss Function}\label{subsec:loss-function}
Prior physical knowledge is integrated with the NN architecture to make it informed; see Fig. \ref{fig:framework}-B. A part of the topology information of the power system, i.e., the diagonal elements of the admittance matrix $Y_{kk}$, is used to develop the physical model. In addition, the real and imaginary components of the complex voltage at the reference bus, i.e., $\mu_0=1, \omega_0=0$, are imported. Eventually, the physical model $f' (\cdot)$ is derived from the prior physical knowledge of the power system. 

The derivation starts with Ohm's law, which relates voltage $V$, current $I$, and resistance $R$, that is, $V=I\times R$. By extending this equation to the complex domain, $\overline{V}=\overline{I}\times\overline{Z}$ relates the voltage phasor, current phasor, and impedance phasor. From this equation, $\overline{I}$ can be expressed in terms of the voltage phasor and admittance phasor $\overline{Y}$ (inverse of the impedance phasor), that is $\overline{I}=\overline{Y}\times\overline{V}$. The system of equations for all the buses can then be presented in matrix form as: 
\begin{align} 
    \begin{bmatrix} I \end{bmatrix}_{n \times 1}=\begin{bmatrix} Y \end{bmatrix}_{n \times n} \times {\begin{bmatrix} V \end{bmatrix}_{n \times 1} }. \label{eq:ohm-law-matrix}
\end{align}
By rearranging \eqref{eq:ohm-law-matrix}, the current flowing into bus $k$ can be presented as a linear combination of the voltages at all other buses with weights given by the corresponding admittance~matrix: 
\begin{align} 
    I_k = \sum_{i=1}^n Y_{ki} V_{i}. \label{eq:I_k}
\end{align}
Here, the power equation $S=V\times I^{*}$ provides further guides for subsequent derivations, that is $I= {S^{*}\over{V^{*}}}$, where $S$ is complex apparent power. The current and voltage phasors are related~as: 
\begin{align} 
    {{S_k^*} \over {V_k^*}} = Y_{k1}V_1 + Y_{k2}V_2 + ... + Y_{kk}V_k + ... + Y_{kn}V_{n}, \label{eq:S_k-V_k}
\end{align}
wherein the right-hand side is an extended form of the right-hand expression in \eqref{eq:I_k}. Finally, the expression for calculating the voltage phasor at bus $k$ can be written as:
\begin{align} 
    V_k = {{1} \over {Y_{kk}}} \times \left( {{S_k^*} \over {V_k^*}} - \sum_{i=1, i \neq k}^n Y_{ki} V_i \right). \label{eq:V_k}
\end{align}
In \eqref{eq:V_k}, $Y_{kk}=G_{kk}+jB_{kk}$ is known from the power system topology. Considering the input features and output labels needed to develop the NN for PF analysis, $V_k=\mu_k+j\omega_k$ and $V_k^*=\mu_k-j\omega_k$ are known from the output labels, and $S_k^*=p_{k}^d+jq_{k}
^d$ is known from the input features. The only unknown expression remaining in \eqref{eq:V_k} is:
\begin{align} 
    \psi_k = \sum_{i=1, i \neq k}^n Y_{ki} V_i, \label{eq:psi_k}
\end{align}
which is called \textit{hidden function} in this study. Note that $\psi$ is unique for each data point. Theoretically, $\psi$ is also unique for each bus. During the training process, the NN first approximates $\psi$ and then uses it to approximate $V=\mu+j\omega$~using:
\begin{align} 
    f'(\Vec{x}) = {{1} \over {Y_{kk}}} \times \left( {{p^d-jq^d} \over {\mu-j\omega}} - \psi \right). \label{eq:physical-model}
\end{align}
This unique relation improves the learning process by following the underlying physical laws of the power system. Note that instead of integrating the whole admittance matrix $\begin{bmatrix} Y \end{bmatrix}_{n \times n}$, only the diagonal elements $Y_{kk}$ are needed to obtain the physical loss term, and therefore, the computational resource required is reduced by up to three orders of magnitude compared to when integrating the whole topology information (e.g., \cite{Yan2025}). Typically, the goal of the training process is to minimize the difference between $\hat{\vec{y}}$, approximated by the NN, and ground-truth output labels $\vec{y}$, obtained from NR, based on a loss function of choice. Here, a modified loss function is developed that integrates both supervised and physical penalty terms. The former is the mean square of the difference between the output approximated by the NN $f(\cdot)$ and $\vec{y}$. While the latter is the mean square of the difference between the output obtained from the physical model $f' (\cdot)$ and $\vec{y}$, as:
\begin{align}
\begin{split}
\mathscr{L} = \beta_0 \times {\underbrace {\frac{1}{N}\sum_{j=1}^N \left( f(\vec{p}_{j}^d, \vec{q}_{j}^d, \theta, \vec{\alpha})-\vec{y}_j\right)^2}_{\mathclap{\text{Supervised penalty term}}} } \\
+ \beta_1 \times {\underbrace {\frac{1}{N}\sum_{j=1}^N \left(f'(\vec{p}_{j}^d, \vec{q}_{j}^d, Y_{kk}, \vec{\psi}_j)-\vec{y}_j\right)^2}_{\mathclap{\text{Physical penalty term}}} },  \label{eq:loss} 
\end{split}
\end{align}
where $N$ is the total number of data points $j$. $f(\cdot) \in \{(f_0, f_1, f_2)\}$ and $f'(\cdot)$ are the NN architecture \eqref{eq:functions-chain-0}--\eqref{eq:functions-chain-2} and the physical model \eqref{eq:physical-model}, respectively. $\beta_0$ and $\beta_1$ represent the coefficients for the supervised and physical penalty terms, respectively, with the constraint that $\beta_0+\beta_1=1$. Initially, $\beta_0 = 1$ and $\beta_1=0$. After 100 epochs, the value of $\beta_1$ increases at a specified rate, with its maximum value determined from the sensitivity analysis. $\theta=\{W,b\}$ is the trainable parameters of the NN and $\alpha$ is the trainable parameter for the activation function. 

\section{Results}
The performance of PINN4PF is evaluated against two baselines: a linear regression model (LR) and a black-box NN (MLP). Experiments are performed on 4-bus \cite{Grainger1994PowerAnalysis}, 15-bus \cite{Farhangi2019MicrogridBenchmarks}, 290-bus \cite{Kerber2011AufnahmefahigkeitPhotovoltaikkleinanlagen}, and 2224-bus \cite{Bukhsh2013NetworkNetworks} test systems. The selected systems contain one reference bus with known voltages in complex number form, $\mu_0 + j\omega_0$, and unknown $p_0^d$ and $q_0^d$, and load buses $i$ for which $p_i^d$ and $q_i^d$ are known, while complex voltages $\mu_i + j\omega_i$ are unknown. 

Input features and output labels of the datasets are represented as $\vec{x}$ and $\vec{y}$, respectively, where $\vec{x}$ includes $(\vec{p}^d_i,\vec{q}^d_i)$ and $\vec{y}$ includes $(\vec{\mu}_i,\vec{\omega}_i)$ obtained from the Newton-Raphson method (NR) for all load buses $i$. The PandaPower Python package \cite{Thurner2018PandapowerAnSystems} is used to perform NR and generate the datasets. Note that PandaPower specifies the state variables of power systems, i.e., $[\delta \text{ } v]^T$, that is, there is a need for converting the state variables to $\mu_i=v_{i}cos\delta_{i}$ and $\omega_i=v_{i}sin\delta_{i}$ to yield the output labels $\vec{y} = \{(\vec{\mu}_i,\vec{\omega}_i): i=1,2,\dots,n\}$.

\subsection{Model Setup}
A systematic approach is employed to generate the datasets. Considering $\vec{p}^d$ and $\vec{q}^d$ known for a specific scenario of the test system, $s_{i}^{d} = \sqrt{ (p_{i}^{d})^2 + (q_{i}^{d})^2}$, and $pf_{i}={p_{i}^{d}/s_{i}^{d}}$ are computed for each bus $i$. Here, $s_{i}^{d}$ is the mean, and a deviation of $30\%$ from $s_{i}^{d}$ is the standard deviation to develop a normal distribution of size 5000 for each bus $i$, i.e., $S_i \sim \mathcal{N}(s_{i}^{d}, 0.3)$. For $S_i \in \{s_{ij}^{d}: j=1,2,\dots,5000 \}$, $p_{ij}^{d}=s_{ij}^{d} \times pf_{i}$ and $q_{ij}^{d}=\sqrt{(s_{ij}^{d})^2 - (p_{ij}^{d})^2}$ are then computed for all buses $i$ and all samples $j$. 

This approach results in a pool of 5000 scenarios from which the data points are randomly selected as the dataset. The datasets contain different numbers of data points: 256, 512, 1024, and 2048 for the 4-bus, 15-bus, 290-bus, and 2224-bus test systems, respectively. Note that the number of data points is deliberately limited to show the superiority of PINN4PF compared to the state-of-the-art deep learning-based approaches, e.g., NNs. Accordingly, the randomly selected data points out of 5000 generated scenarios are sent to the classical PF solver, i.e., PandaPower, to perform PF analysis. Each dataset is then split into three subsets: 40\% for training, 20\% for validation, and the remaining 40\% for testing.

The training process ends after 5000 epochs for PINN4PF, MLP, and LR. The loss function used is \eqref{eq:loss}, with $\beta_0=1$ and $\beta_1=0$ for MLP and LR, indicating a supervised penalty term. However, $\beta_0$ and $\beta_1$ are non-zero for PINN4PF where $\beta_0+\beta_1=1$, indicating a combination of supervised and physical penalty terms. The activation function is ReLU \eqref{eq:relu} for MLP and the shared hidden layers of PINN4PF. The adaptive ReLU \eqref{eq:adaptive-relu} is used for the separated hidden layers of PINN4PF. The Adam optimization algorithm updates the trainable parameters during the training process for PINN4PF, MLP, and LR.

\subsection{Model Performance}
The performance of PINN4PF, MLP, and LR is systematically evaluated based on (i) generalization ability, (ii) robustness, (iii) impact of training dataset size on generalization ability, (iv) accuracy in approximating derived PF quantities, and (v) scalability. Experiments are done using the 15-bus test system. Additional experiments are also performed using 4-bus, 290-bus, and 2224-bus test systems for scalability.

\subsubsection{Generalization ability} 
\begin{figure}[t]
\centering
\includegraphics[width=2.9in]{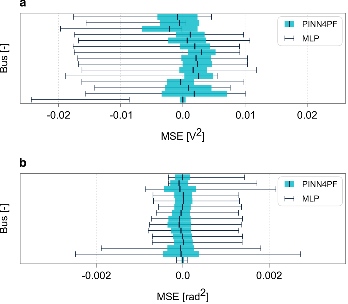}
\caption{Mean and standard deviation of MSE for direct physical quantities: (a) voltage magnitude $[V^2]$ and (b) voltage phase angle $[rad^2]$. The results are obtained by PINN4PF and MLP based on the testing dataset for the 15-bus test system.}
\label{fig:15vd}
\end{figure}

The mean and standard deviation of the mean squared error (MSE) for the test dataset obtained by PINN4PF and MLP are compared in Fig. \ref{fig:15vd}. PINN4PF is observed to achieve up to 85\% and 65\% lower maximum testing MSE for the voltage magnitude $[V^2]$ and voltage phase angle $[rad^2]$, respectively, compared to MLP.

\subsubsection{Robustness} 
\begin{figure}[t]
\centering
\includegraphics[width=2.8in]{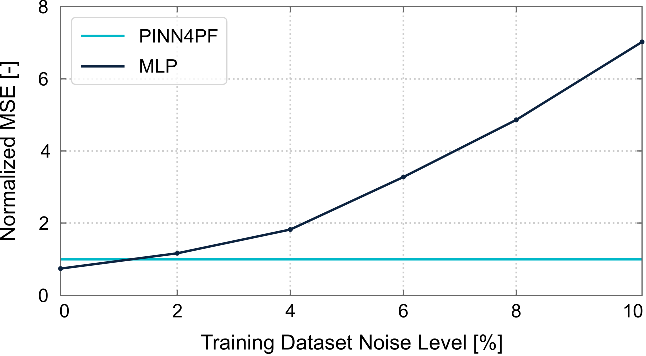}
\caption{Illustration of the performance of PINN4PF and MLP based on the MSE obtained for the voltage magnitude $[V^2]$ under varying noise levels in the training dataset for the 15-bus test system. The MSE values are normalized relative to that of PINN4PF at the 10\% noise level.}
\label{fig:robust}
\end{figure}

The robustness of PINN4PF, MLP, and LR is evaluated using the 15-bus test system under varying noise levels. Controlled Gaussian noise with standard deviations corresponding to noise levels from 0\% to 10\% is added to both input features and output labels of the training dataset. The corrupted vectors are defined as $\vec{x}' = \vec{x} \pm \vec{r}_x$ and $\vec{y}' = \vec{y} \pm \vec{r}_y$, where $\vec{r}_x$ and $\vec{r}_y$ are vectors of random values between 0 and 1, and 0 and 0.1, respectively. Fig. \ref{fig:robust} illustrates the changes in the MSE for the voltage magnitude $[V^2]$ based on the testing dataset with varying noise levels for MLP. The comparison is made relative to a constant curve, representing the MSE obtained for PINN4PF using the highest noise level, i.e., 10\%. At the 0\% noise level, MLP outperforms PINN4PF trained with 10\% noisy data by approximately 18\%. However, as the noise level increases, the performance of MLP deteriorates significantly, with MSE increasing up to six times. LR is excluded from the graph as its MSE at different noise levels is two orders of magnitude higher than that of PINN4PF. In addition, LR shows limited improvement with increasing dataset sizes and consistently underperforms compared to both PINN4PF and MLP.

\subsubsection{Training dataset size} 
\begin{figure}[t]
\centering
\includegraphics[width=2.8in]{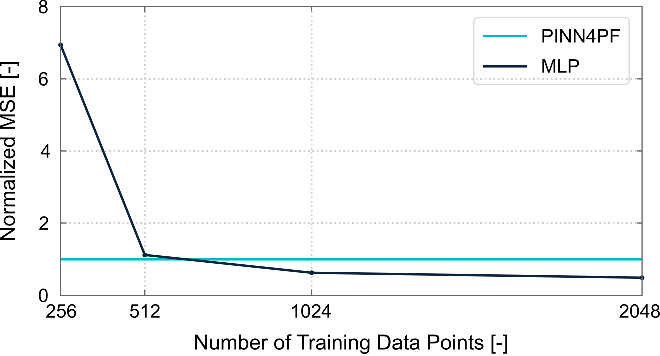}
\caption{Illustration of the performance of PINN4PF and MLP based on the MSE obtained for the voltage magnitude $[V^2]$ for the 15-bus test system under different training dataset sizes. The MSE values are normalized relative to that of PINN4PF with 256 training data points.}
\label{fig:trainsize}
\end{figure}

The impact of the size of the training dataset on the performance of MLP is investigated using the 15-bus test system. Fig. \ref{fig:trainsize} illustrates the changes in the MSE for the voltage magnitude $[V^2]$ based on the testing dataset with varying training dataset sizes. The comparison is made relative to a constant curve representing the MSE obtained for PINN4PF using 256 training data points. It is observed that MLP requires a training dataset twice as large as that of PINN4PF to achieve a still inferior performance. However, the performance of MLP improves with more training data, reaching a comparable level with a dataset four times larger than that of PINN4PF. This indicates that PINN4PF is more data-efficient. LR is not included in the graph as its MSE with different training dataset sizes is two orders of magnitude larger than that of PINN4PF. In addition, LR shows limited improvement with increasing training dataset sizes and consistently performs worse than PINN4PF and MLP. This indicates that the capacity of LR to capture complex relationships in PF analysis is significantly lower, and increasing the amount of training data is insufficient to bridge the performance gap.

\subsubsection{Accuracy of derived power flow quantities} 
\begin{figure}[t]
\centering
\includegraphics[width=2.9in]{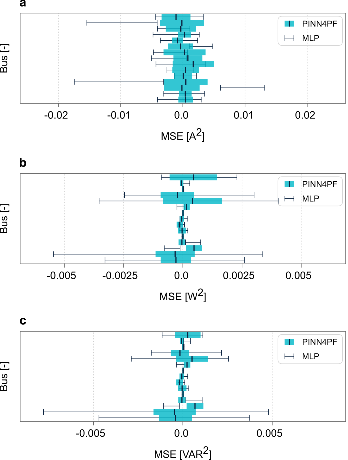}
\caption{Mean and standard deviation of MSE for derived physical quantities: (a) line current $[A^2]$, (b) line active power $[W^2]$, and (c) line
reactive power $[VAR^2]$. The results are obtained by PINN4PF and MLP based on the testing dataset for the 15-bus test system.}
\label{fig:15ipq}
\end{figure}

For derived physical quantities, i.e., line current, line active power, and line reactive power, the mean and standard deviation of the testing MSE obtained by PINN4PF and MLP are computed and compared in Fig. \ref{fig:15ipq}. It is also observed that PINN4PF achieves up to 81\%, 63\%, and 66\% lower maximum testing MSE for the line current $[A^2]$, line active power $[W^2]$, and line reactive power $[VAR^2]$, respectively, compared to MLP.

\subsubsection{Scalability} 
\begin{figure}[t]
\centering
\includegraphics[width=3.1in]{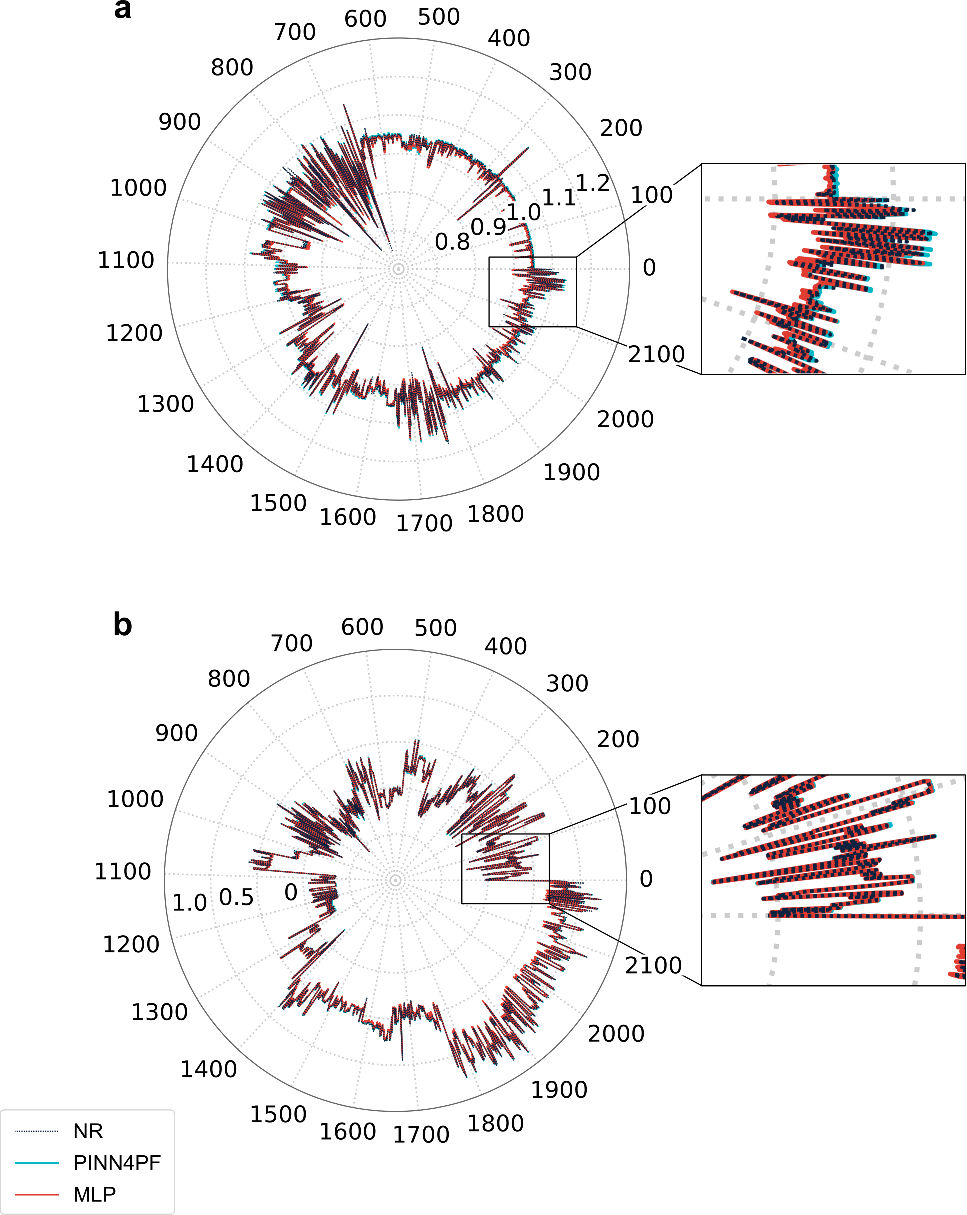}
\caption{Comparison of (a) $\vec{v}$ and (b) $\vec{\delta}$ obtained from the NR and $\vec{\hat{v}}$ and $\vec{\hat{\delta}}$ obtained from PINN4PF and MLP for the 2224-bus test system under extreme conditions.}
\label{fig:2224vd}
\end{figure}

\begin{figure}[t]
\centering
\includegraphics[width=3.1in]{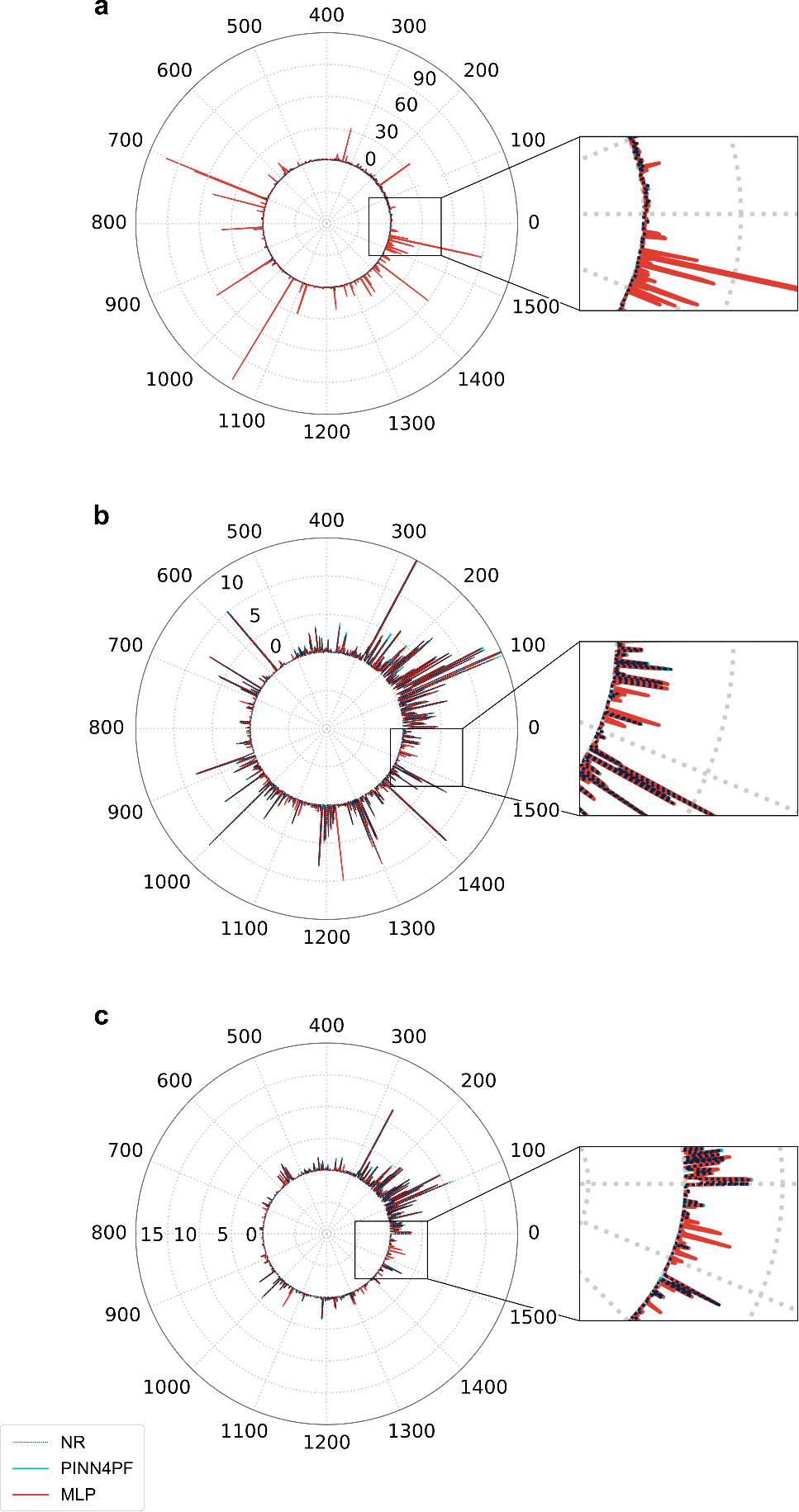}
\caption{Comparison of (a) line current $\vec{i}$, (b) line active power $\vec{p_l}$, and (c) line reactive power $\vec{q_l}$ obtained from NR and $\vec{\hat{i}}$, $\vec{\hat{pl}}$, and $\vec{\hat{ql}}$ obtained from PINN4PF and MLP for the 2224-bus test system under extreme conditions.}
\label{fig:2224ipq}
\end{figure}

The experiments are extended to compare the performance of PINN4PF, MLP, and LR across different test system sizes: 4-bus, 15-bus, 290-bus, and 2224-bus test systems. The maximum testing MSE for the direct physical quantities obtained for all test system sizes is presented in Table II. Accordingly, PINN4PF significantly outperforms MLP and LR. Fig. \ref{fig:2224vd} compares the performance of PINN4PF and MLP in terms of direct physical quantities under extreme conditions for the 2224-bus test system. PINN4PF achieves a maximum MSE $[V^2]$ 50\% lower than MLP for this test system. This improvement is crucial for power system operations, particularly during unexpected events that cause deviations in $v$ from the nominal value. The analysis also considers the performance of PINN4PF, MLP, and LR in approximating derived physical quantities. The maximum testing MSE for all test system sizes is presented in Table III. The table demonstrates the superiority of PINN4PF over MLP and LR across all test system sizes. For the 2224-bus test system, for example, the maximum MSE $[A^2]$ obtained by PINN4PF is respectively 92\% and 98\% lower than MLP and LR. In addition, as the size of the test system increases, the performance gap between PINN4PF, MLP, and LR becomes more pronounced. Fig. \ref{fig:2224ipq} compares the performance of PINN4PF and MLP under extreme conditions in the 2224-bus test system for the derived physical quantity. It should be noted that while the testing MSE for the direct quantities, i.e., voltage magnitude and phase angle, shows marginal differences between PINN4PF and MLP, PINN4PF significantly outperforms MLP in terms of the derived physical quantities, such as line current.

\begin{table}[t] 
\caption{Performance comparison of power flow solvers based on MSE obtained for direct physical quantities.}
\label{tab:vd}
\centering
\footnotesize 
\renewcommand{\arraystretch}{1.5}
\begin{tabularx}{0.4\textwidth}{ >{\raggedright\arraybackslash}p{1cm} >{\raggedright\arraybackslash}p{1cm}  >{\centering\arraybackslash}p{1.5cm} >{\centering\arraybackslash}p{1.5cm}}

\toprule
\scriptsize{Case} & \scriptsize{Model} & \scriptsize{MSE [$V^2$]} & \scriptsize{MSE [$rad^2$]}\\[1pt]
\midrule

\scriptsize{4-bus}    & \scriptsize{PINN4PF} & \tiny{$4.85 \times 10^{-4}$} & \tiny{$3.81 \times 10^{-8}$}\\[1pt]
                      & \scriptsize{MLP}     & \tiny{$5.70 \times 10^{-3}$} & \tiny{$3.72 \times 10^{-7}$}\\[1pt]
                      & \scriptsize{LR}      & \tiny{$2.52 \times 10^{-2}$} & \tiny{$4.70 \times 10^{-6}$}\\[2pt]

\scriptsize{15-bus}   & \scriptsize{PINN4PF} & \tiny{$5.73 \times 10^{-6}$} & \tiny{$1.06 \times 10^{-7}$}\\[1pt]
                      & \scriptsize{MLP}     & \tiny{$3.96 \times 10^{-5}$} & \tiny{$3.09 \times 10^{-7}$}\\[1pt]
                      & \scriptsize{LR}      & \tiny{$6.32 \times 10^{-4}$} & \tiny{$1.57 \times 10^{-5}$}\\[2pt]
 
\scriptsize{290-bus}  & \scriptsize{PINN4PF} & \tiny{$9.54 \times 10^{-10}$} & \tiny{$1.65 \times 10^{-8}$}\\[1pt]
                      & \scriptsize{MLP}     & \tiny{$3.03 \times 10^{-9}$}  & \tiny{$1.78 \times 10^{-8}$}\\[1pt]
                      & \scriptsize{LR}      & \tiny{$1.03 \times 10^{-7}$}  & \tiny{$2.49 \times 10^{-8}$}\\[2pt]

\scriptsize{2224-bus} & \scriptsize{PINN4PF} & \tiny{$1.03 \times 10^{-4}$} & \tiny{$8.15 \times 10^{-5}$}\\[1pt]
                      & \scriptsize{MLP}     & \tiny{$2.07 \times 10^{-4}$} & \tiny{$7.66 \times 10^{-5}$}\\[1pt]
                      & \scriptsize{LR}      & \tiny{$3.62 \times 10^{0}$}  & \tiny{$2.93 \times 10^{-4}$}\\[2pt]

\bottomrule

\end{tabularx}
\end{table}
\begin{table}[t] 
\caption{Performance comparison of power flow solvers based on MSE obtained for derived physical quantities.}
\label{tab:ipq}
\centering
\footnotesize 
\renewcommand{\arraystretch}{1.5}
\begin{tabularx}{0.52\textwidth}{ >{\raggedright\arraybackslash}p{1cm}  >{\raggedright\arraybackslash}p{1cm}  >{\centering\arraybackslash}p{1.5cm}  >{\centering\arraybackslash}p{1.5cm}  >{\centering\arraybackslash}p{1.6cm}}

\toprule
\scriptsize{Case} & \scriptsize{Model} &  \scriptsize{MSE [$A^2$]} & \scriptsize{MSE [$W^2$}] & \scriptsize{MSE [$VAR^2$]}\\[1pt]
\midrule

\scriptsize{4-bus} & \scriptsize{PINN4PF}    & \tiny{$2.43 \times 10^{-6}$} & \tiny{$4.25 \times 10^{-4}$} & \tiny{$1.06 \times 10^{-4}$}\\[1pt]
                   & \scriptsize{MLP}        & \tiny{$2.80 \times 10^{-5}$} & \tiny{$8.27 \times 10^{-3}$} & \tiny{$2.06 \times 10^{-3}$}\\[1pt]
                   & \scriptsize{LR}         & \tiny{$4.04 \times 10^{-5}$} & \tiny{$1.34 \times 10^{-2}$} & \tiny{$3.36 \times 10^{-2}$}\\[2pt]

\scriptsize{15-bus} & \scriptsize{PINN4PF}   & \tiny{$4.35 \times 10^{-6}$} & \tiny{$2.09 \times 10^{-7}$} & \tiny{$2.17 \times 10^{-7}$}\\[1pt]
                    & \scriptsize{MLP}       & \tiny{$2.39 \times 10^{-5}$} & \tiny{$5.77 \times 10^{-7}$} & \tiny{$6.42 \times 10^{-7}$}\\[1pt]
                    & \scriptsize{LR}        & \tiny{$1.10 \times 10^{-2}$} & \tiny{$8.70 \times 10^{-4}$} & \tiny{$1.74 \times 10^{-3}$}\\[2pt]

\scriptsize{290-bus} & \scriptsize{PINN4PF}  & \tiny{$3.84 \times 10^{-7}$} & \tiny{$2.01 \times 10^{-13}$} & \tiny{$4.77 \times 10^{-14}$}\\[1pt]
                     & \scriptsize{MLP}      & \tiny{$2.85 \times 10^{-6}$} & \tiny{$1.10 \times 10^{-12}$} & \tiny{$2.44 \times 10^{-13}$}\\[1pt]
                     & \scriptsize{LR}       & \tiny{$1.36 \times 10^{-5}$} & \tiny{$8.20 \times 10^{-12}$} & \tiny{$1.84 \times 10^{-12}$}\\[2pt]

\scriptsize{2224-bus} & \scriptsize{PINN4PF} & \tiny{$1.15 \times 10^{-6}$} & \tiny{$1.13 \times 10^{-8}$}  & \tiny{$9.45 \times 10^{-7}$}\\[1pt]
                      & \scriptsize{MLP}     & \tiny{$3.69 \times 10^{-5}$} & \tiny{$7.22 \times 10^{-8}$}  & \tiny{$5.12 \times 10^{-5}$}\\[1pt]
                      & \scriptsize{LR}      & \tiny{$7.85 \times 10^{-1}$} & \tiny{$3.55 \times 10^{-1}$}  & \tiny{$2.44 \times 10^{2}$}\\[2pt]

\bottomrule

\end{tabularx}
\end{table}

\subsection{Model Sensitivity Analysis}
The architecture of PINN4PF and MLP, including the number of hidden layers and neurons per hidden layer, as well as hyperparameters, including the learning rate, the weight decay rate, and the percentage of dropout, are determined by sensitivity analysis to ensure a fair comparison\footnote{An exhaustive search across 2000 unique hyperparameter combinations is conducted using the Optuna Python package \cite{akiba2019optuna}.}. The maximum weight for the share of the supervised penalty term ($\beta_0$) in \eqref{eq:loss} is also fine-tuned for PINN4PF. It should be noted that for MLP, $\beta_0=1$ and $\beta_1=0$. Sensitivity analysis is not performed for LR since it involves linear combinations and lacks hyperparameters. The optimal configurations are identified based on minimizing both training and testing MSE, but also ensuring that they remain close enough to indicate generalization ability. All experiments use the 15-bus test system and a noisy dataset comprising $512$ data points.

\subsubsection{Sensitivity analysis for MLP} We explore eight configurations, intentionally avoiding deeper networks with more than eight hidden layers due to the limited size of both the dataset and the test system. Each configuration comprises input and output layers defined by $n \times 2$ neurons in the first and last hidden layers, respectively, while intermediate layers are structured with $n \times 4/3$ neurons. We observe that when the number of hidden layers is fewer than seven, a reduction in layer count correlates with a decrease in generalization ability. We examine learning rates ranging from $1 \times 10^{-1}$ to $1 \times 10^{-10}$ and weight decay rates within the same span. Dropout rates are tested between $0\%$ and $2\%$. Sensitivity analysis results indicate that the selected configuration for MLP involves seven hidden layers and is trained with a learning rate of $2.3 \times 10^{-4}$, a weight decay of $1.8 \times 10^{-5}$, a dropout percentage of $0.2\%$, and a batch size of $16$. For this configuration, the training and testing MSE achieved are $3.74 \times 10^{-5} [V^2]$ and $3.96 \times 10^{-5} [V^2]$, respectively.

\subsubsection{Sensitivity analysis for PINN4PF} We systematically assess the influence of the number of hidden layers, ranging from one to four, within the shared set of hidden layers and the individual heads. We evaluate distinct configurations where the shared hidden layers contain $n \times 2$ neurons, and each head comprises $n$ neurons. Learning and weight decay rates are varied from $1 \times 10^{-1}$ to $1 \times 10^{-10}$, and dropout rates are varied from $0\%$ to $2\%$. Sensitivity analysis reveals that the selected configuration for PINN4PF consists of two shared hidden layers alongside four hidden layers per head. In addition, PINN4PF is trained with a learning rate of $1.3 \times 10^{-4}$, a weight decay of $1.1 \times 10^{-5}$, a dropout rate of $0.1\%$, and a batch size of $16$. The maximum weight for the share of the supervised penalty term is $\beta_1 = 0.71$. For this configuration, the achieved training and testing MSE are $5.64 \times 10^{-6} [V^2]$ and $5.73 \times 10^{-6} [V^2]$, respectively.

\subsubsection{Impact of modifications on the performance of PINN4P} We evaluate the following configurations: (i) a double-head feed-forward NN with ReLU \eqref{eq:relu} and supervised penalty term (ReLU\&Supervised), (ii) a double-head feed-forward NN with adaptive ReLU \eqref{eq:adaptive-relu} and supervised penalty term (AdaptiveReLU\&Supervised), (iii) a double-head feed-forward NN with ReLU \eqref{eq:relu} and physical penalty term (ReLU\&Physical), and (iv) a double-head feed-forward NN with adaptive ReLU \eqref{eq:adaptive-relu} and physical penalty term (AdaptiveReLU\&Physical) that is the selected configuration for PINN4PF. Table IV presents a comparative analysis of their performance in terms of direct physical quantities. The results show that combining the physics-informed loss function with the adaptive activation function yields superior performance of PINN4PF than the other combinations.

\begin{table}[t] 
\caption{Performance comparison of different configurations of PINN4PF based on MSE obtained for direct physical quantities and the 15-bus test system.}
\label{tab:vd}
\centering
\footnotesize 
\renewcommand{\arraystretch}{1.5}
\begin{tabularx}{0.45\textwidth}{ > {\raggedright\arraybackslash}p{3.2cm}  >{\centering\arraybackslash}p{1.6cm}>
{\centering\arraybackslash}p{1.6cm} }

\toprule
\scriptsize{Model} & \scriptsize{MSE [$V^2$]} & \scriptsize{MSE [$rad^2$]} \\ [1pt]
\midrule

\scriptsize{ReLU\&Supervised} & \tiny{$4.05 \times 10^{-5}$} & \tiny{$3.24 \times 10^{-7}$} \\[1pt]
\scriptsize{AdaptiveReLU\&Supervised} &  \tiny{$1.42 \times 10^{-5}$} & \tiny{$3.73 \times 10^{-7}$}  \\[1pt]
\scriptsize{ReLU\&Physical} &  \tiny{$6.31 \times 10^{-6}$} & \tiny{$4.70 \times 10^{-7}$} \\[2pt]
\scriptsize{AdaptiveReLU\&Physical} &  \tiny{$5.73 \times 10^{-6}$} & \tiny{$1.06 \times 10^{-7}$} \\[2pt]

\bottomrule
\end{tabularx}
\end{table}

\section{Discussion}
Several key points and observations arise from the findings:
\begin{itemize}
    \item The optimization algorithm suggests that for larger test systems, the share of the supervised penalty term ($\beta_0$) in the loss function should be significantly smaller than the physical penalty term ($\beta_1$). This indicates the increasing importance of physical constraints in larger test systems. 
    \item Although only the diagonal elements of the admittance matrix are used to develop the loss function, including a physical penalty term imposes additional calculations during training, which makes PINN4PF computationally more expensive than supervised models and highlights a trade-off between accuracy and computational cost. 
    \item Additional testing on real-world power systems and incorporating other physical constraints could further validate and improve the model's performance.
    \item PINN4PF is trained on different operating conditions, and hence, can generalize to provide approximate solutions even in ill-conditioned scenarios, where traditional methods fail to converge. Future work should explore the use of these approximate solutions as initial guesses (warm starts) for traditional methods and assess their potential to enhance convergence in such challenging cases.
    \item In modern power systems, missing or noisy measurements and incomplete topology data are common, often causing traditional PF solvers to fail. PINN4PF addresses these challenges by integrating partial topology information, specifically, the diagonal elements of the admittance matrix, alongside combining supervised and physics-based penalty terms. This design enables PINN4PF to learn complex system dynamics from limited data and enhances robustness against noise or incompleteness in both training samples and topology information. Future work may further assess PINN4PF’s resilience against noisy data in the admittance matrix.
    \item While PINN4PF demonstrates strong performance under fixed network topology, it requires retraining if the system topology changes, as the current architecture does not support topological generalization. Future research may explore the potential of transfer learning to adapt pre-trained models to new topologies with minimal additional training. Note, however, that although recent advances in GNNs aim to support topological changes, they still do not provide consistently reliable results at the time of writing this paper. Therefore, enabling robust topological adaptability remains an open research challenge for NNs.
    \end{itemize}

\section{Conclusion}
This study presents an end-to-end architecture for deep learning-based power flow (PF) analysis called PINN4PF. PINN4PF includes three significant modifications: a double-head feed-forward neural network, an adaptive activation function, and a physics-based loss function. PINN4PF, therefore, offers a straightforward yet effective approach to capturing the complexities inherent in large-scale modern power systems. The application of PINN4PF to four test power systems, including 4-bus, 15-bus, 290-bus, and 2224-bus systems, rigorously evaluates its performance against two baselines: a linear regression model (LR) and a black-box neural network (MLP). 

The results highlight PINN4PF's exceptional generalization ability, robustness against noise, data efficiency, and scalability to large-scale power systems. Specifically, PINN4PF consistently achieves lower mean squared errors for direct and derived physical quantities, proving its effectiveness and reliability. In addition, as the test system size increases, the performance gap between PINN4PF, MLP, and LR becomes more pronounced. This advancement is crucial for enhancing power system operations, especially under extreme conditions and unexpected deviations, making PINN4PF a promising solution for future PF analysis and optimization tasks. 

\section*{Acknowledgment}
This work is part of the DATALESs project (project no. 482.20.602), which is jointly financed by Netherlands Organization for Scientific Research (NWO), and National Natural Science Foundation of China (NSFC). The work used the Dutch national e-infrastructure with the support of the SURF Cooperative (grant no. EINF-6569).

\bibliography{main}

\end{document}